\documentclass[12pt]{iopart}
\usepackage{epsfig}
\usepackage{citesort}

\newcommand{\gtrsim}{\,\rlap{\lower3.7pt\hbox{$\mathchar\sim$}}
\raise1pt\hbox{$>$}\,}
\newcommand{\lesssim}{\,\rlap{\lower3.7pt\hbox{$\mathchar\sim$}}
\raise1pt\hbox{$<$}\,}

\begin{document}

\title{New constraint on the cosmological background of relativistic particles}
\author{Steen Hannestad}
\address{Department of Physics and Astronomy,
University of Aarhus,
Ny Munkegade, DK-8000 Aarhus C, Denmark\\}
\date{\today}

\begin{abstract}
We have derived new bounds on the relativistic energy density in
the Universe from cosmic microwave background (CMB), large scale
structure (LSS), and type Ia supernova (SNI-a) observations. In
terms of the effective number of neutrino species a bound of
$N_\nu = 4.2^{+1.2}_{-1.7}$ is derived at 95\% confidence. This
bound is significantly stronger than previous determinations,
mainly due to inclusion of new CMB and SNI-a observations. The
absence of a cosmological neutrino background ($N_\nu = 0$) is now
excluded at $5.4 \sigma$. The value of $N_\nu$ is compatible with
the value derived from big bang nucleosynthesis considerations,
marking one of the most remarkable successes of the standard
cosmological model. In terms of the cosmological helium abundance,
the CMB, LSS, and SNI-a observations predict a value of $0.240 < Y
< 0.281$.
\end{abstract}

\maketitle

\section{Introduction} 

From detailed observations of the cosmic microwave background
(CMB)
\cite{Bennett:2003bz,Spergel:2003cb,Verde:2003ey,Kogut:2003et,%
Hinshaw:2003ex,Jones:2005yb,Piacentini:2005yq,Montroy:2005yx}, the
large scale structure (LSS) of galaxies
\cite{Tegmark:2003ud,Tegmark:2003uf,2dFGRS}, and distant type Ia
supernovae (SNI-a)
\cite{Riess:1998cb,Perlmutter:1998np,Riess:2004} many of the
fundamental cosmological parameters are known quite precisely.
Very interestingly the precision of the data is now also at a
level where particle physics beyond the standard model can be
probed. One of the prime examples of this is the total
cosmological energy density in relativistic particles.

In the standard model photons and neutrinos are by far the largest
source of entropy in the Universe (and consequently also of energy
density in the early Universe), and neutrinos are therefore the
largest source of entropy in non-electromagnetically interacting
species.

In the early Universe neutrinos decouple at a temperature of
roughly 2-3 MeV. Shortly after this, electrons and positrons
annihilate and transfer entropy to the photons. The end result is
that the temperature of neutrinos is roughly $(4/11)^{1/3}
T_\gamma$. The total energy density in weakly interacting
relativistic species is therefore $\rho_\nu \sim 3 (4/11)^{4/3}
\rho_\gamma \sim 0.78 \rho_\gamma$.

Any additional relativistic energy density can be thought of as
additional neutrinos, and from the perspective of late time
evolution after neutrino decoupling it is customary to
parameterize any such additional energy density in terms of
$N_\nu$ \cite{Steigman:kc}, the equivalent number of neutrino
species
\begin{equation}
N_\nu = \frac{\rho}{\rho_{\nu,0}},
\end{equation}
where $\rho_{\nu,0}$ is the density in a massless standard model
neutrino species which has a temperature of exactly
$(4/11)^{1/3}T_\gamma$. The value predicted by the standard model
is actually $N_\nu = 3.04$ because of finite-temperature QED
effects and incomplete neutrino decoupling (see for instance
\cite{dolgov} for a review or \cite{Mangano:2005cc} for a very
recent study).

In general, a species, $X$, decoupling from thermal equilibrium
while relativistic contributes an $N_{\nu,X}$ given by (see for
instance \cite{kolb})
\begin{equation}
N_{\nu,X}=g_X \, \left(\frac{10.75}{g_{*,D}}\right)^{4/3} \times
\cases{1 \,\, ({\rm fermions}) \cr \frac{8}{7} \,\, ({\rm
bosons})},
\end{equation}
where $g_X$ is the number of helicity states of the species and
$g_{*,D}$ is the effective number of degrees of freedom in the
plasma at the time of $X$-decoupling.

In the present paper we calculate cosmological constraints on
$N_\nu$ from presently available data. In the past the main source
for constraints on $N_\nu$ has been big bang nucleosynthesis
(BBN). However, recently observations of the CMB, LSS, and distant
type Ia supernovae have reached a precision where they can be used
to constrain $N_\nu$ at a competitive level
\cite{Hannestad:2000hc,Hannestad:2001hn,Hansen:2001hi,Zentner:2001zr,%
Crotty:2003th,Pierpaoli:2003kw,%
Hannestad:2003xv,Barger:2003zg,Barger:2003rt,Cuoco:2003cu,%
Hannestad:2003ye,Crotty:2004gm}.
Here we perform a detailed and updated calculation of CMB, LSS,
and SNI-a constraints on $N_\nu$ and compare them with present BBN
constraints. In Section 2 we discuss present observational
constraints, and section 3 contains a discussion.

\section{Current constraints} 

Using the presently available precision data we have performed a
likelihood analysis for the relativistic energy density,
parameterized in units of the effective number of neutrino
species, $N_\nu$. We assume the relativistic energy density to be
distributed in $N_\nu$ light Majorana fermion species, each with
mass $m_\nu$. The standard case with 3 very light neutrino species
plus possible additional relativistic species just corresponds to
taking $m_\nu \to 0$. The reason why $m_\nu$ is allowed to be
different from 0 is that there is a well-known degeneracy between
$m_\nu$ and $N_\nu$ in CMB and LSS data
\cite{Hannestad:2003xv,Hannestad:2003ye,Crotty:2004gm}.

As our framework we choose a flat dark energy model with the
following free parameters: $\Omega_m$, the matter density, the
curvature parameter, $\Omega_b$, the baryon density, $w$, the dark
energy equation of state, $H_0$, the Hubble parameter, $n_s$, the
spectral index of the initial power spectrum, and $\tau$, the
optical depth to reionization. The normalization of both CMB and
LSS spectra are taken to be free and unrelated parameters. The
dark energy density is given by the flatness condition
$\Omega_{\rm DE} = 1 - \Omega_m - \Omega_\nu$.

The priors we use are given in Table~\ref{table:priors}. The prior
on the Hubble constant comes from the HST Hubble key project value
of $h_0 = 0.72 \pm 0.08$ \cite{freedman}, where $h_0 = H_0/100 \,
{\rm km} \, {\rm s}^{-1} \, {\rm Mpc}^{-1}$.

\begin{table}
\begin{center}
\begin{tabular}{lcl}
\hline \hline parameter & prior\cr
\hline $\Omega=\Omega_m+\Omega_{\rm DE} + \Omega_\nu$&1&Fixed\cr
$h$ & $0.72 \pm 0.08$&Gaussian \cite{freedman}\cr $\Omega_b h^2$ &
0.014--0.040&Top hat\cr $m_\nu$ & 0 -- 5 eV & Top hat \cr $w_{\rm
DE}$ & -2.5 -- -0.5 & Top hat \cr $n_s$ & 0.6--1.4& Top hat\cr
$\tau$ & 0--1 &Top hat\cr $Q$ &
--- &Free\cr $b$ & --- &Free\cr \hline \hline
\end{tabular}
\end{center}
\caption{The different priors on parameters used in the likelihood
analysis.} \label{table:priors}
\end{table}

Likelihoods are calculated from $\chi^2$ so that in 2-dimensional
plots the 68\% and 95\% regions are formally defined by $\Delta
\chi^2 = 2.30$ and 6.17 respectively. Note that this means that
the 68\% and 95\% contours are not necessarily equivalent to the
same confidence level for single parameter estimates. In
1-dimensional plots the same formal confidence levels are given by
$\Delta \chi^2 = 1$ and 4.

\subsection{Type Ia supernovae (SNI-a).}
We perform our likelihood analysis using the ``gold'' dataset
compiled and described in Riess et al \cite{Riess:2004} consisting
of 157 supernovae using a modified version of the SNOC package
\cite{Goobar:2002c}.

\subsection{Large Scale Structure (LSS).}

Any large scale structure survey measures the correlation function
between galaxies. In the linear regime where fluctuations are
Gaussian the fluctuations can be described by the galaxy-galaxy
power spectrum alone, $P(k) = |\delta_{k,gg}|^2$. In general the
the galaxy-galaxy power spectrum is related to the matter power
spectrum via a bias parameter, $b^2 \equiv P_{gg}/P_m$. In the
linear regime, the bias parameter is approximately constant, so up
to a normalization constant $P_{gg}$ does measure the matter power
spectrum.

At present there are two large galaxy surveys of comparable size,
the Sloan Digital Sky Survey (SDSS)
\cite{Tegmark:2003uf,Tegmark:2003ud} and the 2dFGRS (2~degree
Field Galaxy Redshift Survey) \cite{2dFGRS}. Once the SDSS is
completed in 2005 it will be significantly larger and more
accurate than the 2dFGRS. In the present analysis we use data from
both surveys. In the data analysis we use only data points on
scales larger than $k = 0.15 h$/Mpc in order to avoid problems
with non-linearity.

\subsection{Cosmic Microwave Background (CMB).}

The temperature fluctuations are conveniently described in terms
of the spherical harmonics power spectrum $C_{T,l} \equiv \langle
|a_{lm}|^2 \rangle$, where $\frac{\Delta T}{T} (\theta,\phi) =
\sum_{lm} a_{lm}Y_{lm}(\theta,\phi)$.  Since Thomson scattering
polarizes light, there are also power spectra coming from the
polarization. The polarization can be divided into a curl-free
$(E)$ and a curl $(B)$ component, yielding four independent power
spectra: $C_{T,l}$, $C_{E,l}$, $C_{B,l}$, and the $T$-$E$
cross-correlation $C_{TE,l}$.

The WMAP experiment has reported data on $C_{T,l}$ and $C_{TE,l}$
as described in
Refs.~\cite{Spergel:2003cb,Bennett:2003bz,Kogut:2003et,%
Hinshaw:2003ex,Verde:2003ey}.  We have performed our likelihood
analysis using the prescription given by the WMAP
collaboration~\cite{Spergel:2003cb,Bennett:2003bz,Kogut:2003et,%
Hinshaw:2003ex,Verde:2003ey} which includes the correlation
between different $C_l$'s. Foreground contamination has already
been subtracted from their published data.

We furthermore use the newly published results from the Boomerang
experiment \cite{Jones:2005yb,Piacentini:2005yq,Montroy:2005yx}
which has measured significantly smaller scales than WMAP.

\subsection{Results}

\begin{figure}[htb]
\begin{center}
\epsfysize=8truecm\epsfbox{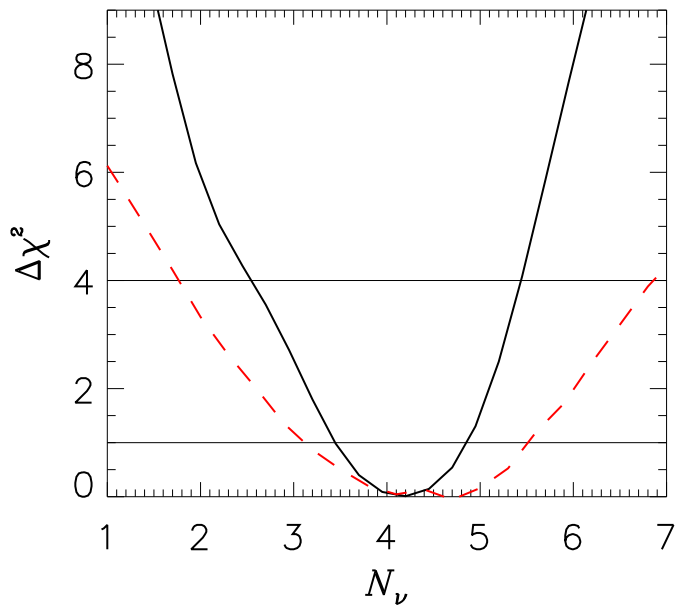}
\end{center}
\caption{$\Delta \chi^2$ values as a function of $N_\nu$ for
various data sets. The full line includes all available data, and
the dashed line is for WMAP and LSS data only.} \label{fig1}
\end{figure}

In Fig.~\ref{fig1} we show the results of the likelihood analysis
for all the currently available data. As with previous analyses
which include WMAP data the best fit is higher than $N_\nu = 3$,
but with $N_\nu = 3$ well within the 95\% contour. In terms of
$N_\nu$ the formal allowed region is
\begin{equation}
N_\nu = 4.2^{+1.2}_{-1.7}  \,\, (95\% \, {\rm C.L.}).
\end{equation}
This can be compared with the previous limit of
\cite{Crotty:2004gm}
\begin{equation}
1.6 < N_\nu < 7.2 \,\, (95\% \, {\rm C.L.}).
\end{equation}
It is quite interesting to compare these limits. The parameter
space used in the present analysis is larger (taking $w$ as a free
parameter), but also uses newer data (the Riess et al. SNI-a data
and the Boomerang data). The new data significantly strengthens
the constraint and clearly the severe degeneracy between $w$ and
$m_\nu$ which affects the neutrino mass bound
\cite{Hannestad:2005gj} has no significant impact on the
determination of $N_\nu$.

Interestingly the value $N_\nu = 0$ is excluded at $5.4\sigma$,
confirming the presence of a cosmological background of
relativistic particles. However, at present observations are not
precise enough to determine the exact nature of the relativistic
background, such as whether it is of fermionic or bosonic nature.
Other studies have shown that the relativistic background must be
weakly interacting around the epoch of recombination
\cite{Hannestad:2004qu,Trotta:2004ty}. At present all observations
are therefore compatible with standard model neutrinos as the only
non-electromagnetically interacting relativistic background, but
cannot definitively rule out a background of other light
particles.

In order to compare with the previous analysis, we also show
$\chi^2$ for an analysis with only WMAP and LSS data, but keeping
$m_\nu = 0$. The 95\% allowed region is in that case $1.7 < N_\nu
< 6.9$, in very good agreement with Fig. 2b in
\cite{Crotty:2004gm}. Fig.~1 shows the quite significant
improvement from including the new Boomerang and SNI-a data.

\subsection{Comparison with Big Bang Nucleosynthesis}

The other main probe of relativistic energy density in the early
Universe is big bang nucleosynthesis (BBN). It is therefore highly
interesting to compare the BBN inferred values of $\Omega_b h^2$
and $N_\nu$ in order to check consistency.

At present the measured primordial helium-4 abundance is quite
uncertain (see for instance \cite{steigman} for a recent
discussion). One of the most recent determinations is
\cite{Olive:2004kq,Cyburt:2004yc}
\begin{equation}
Y = 0.2495 \pm 0.0092,
\end{equation}
where $Y$ is the $^4$He mass fraction. However, the systematic
uncertainty in the measurement is almost certainly larger than
this.

The primordial deuterium fraction is fairly well measured in high
redshift absorption systems. We use the value \cite{Cyburt:2004yc}
\footnote{Note that one very recent study found a significantly
lower value of $D/H = 1.6^{+0.25}_{-0.30} \times 10^{-5}$
\protect\cite{Crighton:2004aj}. Evidently the uncertainty in the
primordial deuterium abundance may also have been
underestimated.}.
\begin{equation}
\frac{D}{H} = (2.78 \pm 0.29) \times 10^{-5}.
\end{equation}

Based on these values we have calculated the allowed region in
$\Omega_b h^2, N_\nu$ space. The result is shown in Fig.~2. The
BBN determination is clearly compatible with the CMB+LSS+SNI-a
determination. Very interestingly the uncertainty in $N_\nu$ from
CMB+LSS+SNI-a is now comparable to what can be obtained from BBN.
Using the $^4$He and D abundances quoted above Cyburt et al.
\cite{Cyburt:2004yc} find that $N_\nu = 3.14^{+0.7}_{-0.65}$ at
68\% C.L. This yields a 95\% bound on $\Delta N_\nu$ of roughly
1.4, comparable to the present bound from CMB, LSS, SNI-a.

\begin{figure}[t]
\vspace*{-0.0cm}
\begin{center}
\epsfysize=8truecm\epsfbox{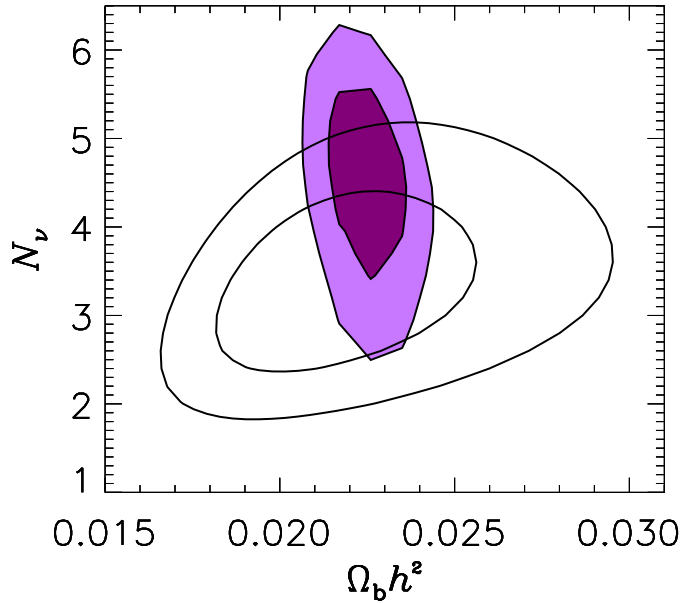} \vspace{0.5truecm}
\end{center}
\caption{The 68\% (dark) and 95\% (light) likelihood contours for
$\Omega_b h^2$ and $N_\nu$ for all available data. The other
contours are 68\% and 95\% regions for BBN, assuming the $^4$He
and D values given in \protect\cite{Cyburt:2004yc}.} \label{fig2}
\end{figure}

\subsection{What are the predictions for primordial $^4$He?}

Given the stringent constraint on $\Omega_b h^2$ and $N_\nu$ from
CMB, LSS, and SNI-a data, it is worthwhile to consider the bound
on the primordial $^4$He value which can be derived from these
data. At 95\% C.L. the bound is
\begin{equation}
0.240 < Y < 0.281.
\end{equation}
This is entirely compatible with the value $Y = 0.2495 \pm 0.0092$
recently derived. However, the allowed region is significantly
larger than the value of $Y = 0.238 \pm 0.005$ suggested a few
years ago based on a large sample of low-metallicity systems
\cite{Olive:1999ij}. In Fig.~3 we show isocontours for $^4$He as a
function of $\Omega_b h^2$ and $N_\nu$ overlayed on the likelihood
analyses.

It should be noted that the allowed region is much larger than the
value quoted in \cite{Cyburt:2004cq}, $Y = 0.2485 \pm 0.0005$ (in
a comparable study \cite{Serpico:2004gx} found $Y = 0.2481 \pm
0.0002 \, [{\rm stat}] \pm 0.0004 \, [{\rm syst}]$, assuming
$\Omega_b h^2 = 0.023$.). The reason is that this assumes a fixed
value of $N_\nu = 3$ and therefore only applies if light, standard
model neutrinos is the only source of relativistic energy.

\begin{figure}[htb]
\begin{center}
\epsfysize=8truecm\epsfbox{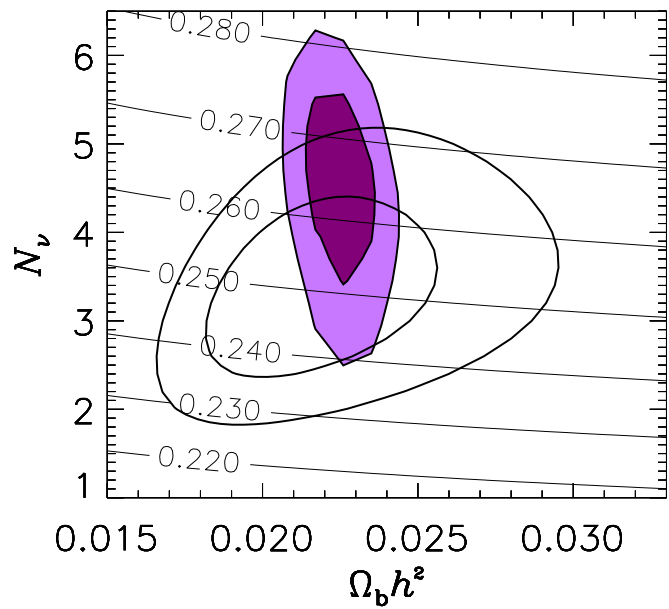}
\end{center}
\caption{Isocontours for $Y$ as a function of $\Omega_b h^2$ and
$N_\nu$. The likelihood contours are as in Fig.~2.} \label{fig3}
\end{figure}

Because the primordial Helium abundance has a direct influence on
recombination it is in principle also possible to measure $Y$ from
CMB data alone. However, at present the bound is not competitive
(\cite{Trotta:2003xg} found a presently allowed range of $0.160 <
Y < 0.501$ at 68\% C.L.). With future data the helium
determination from CMB alone will improve significantly and allow
for one more important consistency check between BBN and CMB.

\section{Discussion}

We have calculated updated constraints on the cosmological
background of relativistic particles from CMB, LSS, and SNI-a. In
the present analysis, a larger set of cosmological parameters has
been used, as well as new CMB and SNI-a data. We find, at the 95\%
C.L. that the bound on the relativistic energy density in terms of
the effective number of neutrino species is $N_\nu =
4.2^{+1.2}_{-1.7}$. The precision of this bound is now comparable
to, or better than, what can be found from big bang
nucleosynthesis considerations. Furthermore, the systematic
uncertainties plaguing the determination of the primordial helium
determination makes any exact BBN bound hard to obtain.

The present data also shows evidence for a cosmological background
of weakly interacting, relativistic energy density at the $5.4
\sigma$ level, an important confirmation of the standard
cosmological model.

The uncertainty in $\Delta N_\nu$ has decreased from roughly 10
five years \cite{Hannestad:2000hc} ago to about 1.5 at present, an
impressive improvement on a very short timescale. Within the next
five years the measurement of $N_\nu$ from CMB data alone could
reach the 0.1 precision level \cite{Lopez:1998aq,Bowen:2001in},
indicating an improvement by two orders of magnitude in the span
of a decade.

\section*{Acknowledgments} 

I wish to thank Pasquale Serpico, Gary Steigman, and Roberto
Trotta for valuable comments.

Use of the publicly available CMBFAST package~\cite{CMBFAST} and
of computing resources at DCSC (Danish Center for Scientific
Computing) are acknowledged.

\clearpage

\section*{References} 

\end{document}